\documentclass[prd,amsmath,amssymb,longbibliography,superscriptaddress,twocolumn,nofootinbib,10pt]{revtex4}

\pdfoutput=1

\usepackage{graphicx}
\usepackage{dcolumn}
\usepackage{bm}
\usepackage{amssymb}
\usepackage{latexsym}
\usepackage{booktabs}
\usepackage{amsmath}
\usepackage{multirow}

\usepackage[colorlinks=true, linkcolor=blue, citecolor=blue]{hyperref}

\begin{document}

\title{Strongly lensed type Ia supernovae as a precise late-universe probe of measuring the Hubble constant and cosmic curvature}

\author{Jing-Zhao Qi}
\affiliation{Department of Physics, College of Sciences, Northeastern University, Shenyang 110819, China}
\author{Yu Cui}
\affiliation{Department of Physics, College of Sciences, Northeastern University, Shenyang 110819, China}
\author{Wei-Hong Hu}
\affiliation{Department of Physics, College of Sciences, Northeastern University, Shenyang 110819, China}
\author{Jing-Fei Zhang}
\affiliation{Department of Physics, College of Sciences, Northeastern University, Shenyang 110819, China}
\author{Jing-Lei Cui}
\affiliation{Department of Physics, College of Sciences, Northeastern University, Shenyang 110819, China}
\author{Xin Zhang\footnote{Corresponding author}}
\email{zhangxin@mail.neu.edu.cn}
\affiliation{Department of Physics, College of Sciences, Northeastern University, Shenyang 110819, China}
\affiliation{Frontiers Science Center for Industrial Intelligence and Systems Optimization, Northeastern University, Shenyang 110819, China}
\affiliation{Key Laboratory of Data Analytics and Optimization for Smart Industry (Northeastern University), Ministry of Education, China}

\begin{abstract}
Strongly lensed type Ia supernovae (SNe Ia) are expected to have some advantages in measuring time delays of multiple images, and so they have a great potential to be developed into a powerful late-universe cosmological probe. In this paper, we simulate a sample of lensed SNe Ia with time-delay measurements in the era of the Legacy Survey of Space and Time (LSST). Based on the distance sum rule, we use lensed SNe Ia to implement cosmological model-independent constraints on the Hubble constant $H_0$ and cosmic curvature parameter $\Omega_K$ in the late universe. We find that if 20 lensed SNe Ia could be observed, the constraint on $H_{0}$ is better than the measurement by the SH0ES collaboration. When the event number of lensed SNe Ia increases to 100, the constraint precision of $H_{0}$ is comparable with the result from \emph{Planck} 2018 data. Considering 200 lensed SNe Ia events as the optimistic estimation, we obtain $\Delta H_0=0.33$ $\rm km\ s^{-1}\ Mpc^{-1}$ and $\Delta\Omega_K=0.053$. In addition, we also simulate lensed quasars in different scenarios to make a comparison and we find that they are still a useful cosmological probe even though the constraint precision from them is much less than that obtained from lensed SNe Ia. In the era of LSST, the measurements of time delay from both lensed SNe Ia and lensed quasars are expected to yield the results of $\Delta H_0=0.26 ~\rm km\ s^{-1}\ Mpc^{-1}$ and $\Delta\Omega_K=0.044$.

\end{abstract}

\maketitle

%\bigskip
\section{Introduction}

The precise measurements of the cosmic microwave background (CMB) anisotropies lead us to an era of precision cosmology \cite{WMAP:2003ivt, WMAP:2003elm}. The $\Lambda$ cold dark matter ($\Lambda$CDM) model with six base parameters, also known as the standard model cosmology, has been constrained by the \emph{Planck}-satellite data with breathtaking precision \citep{Aghanim:2018eyx}. Moreover, in the framework of the flat $\Lambda$CDM model, the constraints from various observations were well consistent with each other \citep{SNLS:2011lii,SupernovaCosmologyProject:2011ycw,2dFGRS:2005yhx,Cao:2014jza,Cao:2015qja,Cao:2017ivt}. However, with the improvements of precision of observations, it was found that inconsistencies emerged between different measurements of some key cosmological parameters. At present, for example, the most perplexing problem is the Hubble constant tension \citep{DiValentino:2021izs,Vagnozzi:2019ezj,Zhang:2019ylr,Qi:2019zdk,Vattis:2019efj,Zhang:2014ifa,Guo:2018ans,Zhao:2017urm,Guo:2017qjt}. Specifically, the CMB power spectra with exquisite precision from the \emph{Planck}-satellite observation, as an early-universe measurement, predict the Hubble constant $H_0$ having a relatively low value of $H_{0}=67.4\pm0.5~\rm{km\ s^{-1}\ Mpc^{-1}}$ assuming a flat $\Lambda$CDM model.  For an end-to-end test, it is necessary to measure the Hubble constant using the late-universe observations, which is implemented by the local type Ia supernovae (SNe Ia) data calibrated by the distance ladder. In this way, the SH0ES (SNe, H0, for the Equation of State of dark energy) collaboration \citep{Riess:2020fzl} reported a high value of the Hubble constant $H_{0}=73.2\pm1.3$ $\rm km\ s^{-1}\ Mpc^{-1}$. Obviously, there is a significant tension with 4.2$\sigma$ disagreement between the values inferred from the two independent methods, which cannot be solely attributed to systematic errors \citep{DiValentino:2018zjj,Riess:2019cxk}.

Recently, some studies \citep{DiValentino:2019qzk, DiValentino:2020hov, Handley:2019tkm} indicated that cosmological tensions are likely to be more severe when the possibility of a closed universe is considered. An enhanced lensing amplitude in CMB power spectra prefers a closed universe at more than $3.4 \sigma$ confidence level \citep{Aghanim:2018eyx,DiValentino:2019qzk}. Moreover, it was found that there are significant tensions between {\it Planck} and low-redshift baryon acoustic oscillations (BAO) data in measuring the curvature parameter $\Omega_K$ \citep{DiValentino:2019qzk,Handley:2019tkm}. All of these mentioned above imply that there exist measurement inconsistencies between the early and late universe in the standard $\Lambda$CDM model. For resolving these cosmological tensions, {it is necessary to} develop novel late-universe observational methods to accurately measure these related cosmological parameters.

%Whether to further confirm these cosmological tensions or to resolve them, there is an urgent need to accurately measure these related parameters in the late universe with new methods as well.

Strong gravitational lensing time delays (SGLTD) as an important probe of the late universe provide a one-step measurement of $H_0$. This method was originally proposed by \citet{Refsdal:1964nw}, in which the time delay is derived from the different arrival times of multiple images generated by the strongly lensed supernovae. However, the lensed SNe events are relatively rare in the universe. Up to now, only two lensed SNe systems have been discovered, namely, iPTF16geu \citep{Goobar:2016uuf} and SN Refsdal \citep{Kelly:2014mwa}. In fact, {the more mature time-delay measurements that have been used in cosmology are from more abundant lensed quasars. Recently,} the H0LiCOW (H0 Lenses in COSMOGRAIL Wellspring) collaboration presented the observations of time delays from 6 lensed quasars, and used them to achieve a 2.4\% precision measurement on $H_0$, i.e., $H_0=73.3^{+1.7}_{-1.8}~\rm{km\ s^{-1}\ Mpc^{-1}}$, in the spatially flat $\Lambda$CDM model \citep{Wong:2019kwg}. Subsequently, the TDCOSMO (Time-Delay COSMOgraphy) collaboration achieved a 2\% precision measurement on $H_0$ with 7 time-delay lensed quasars \citep{Millon:2019slk,Rusu:2019xrq,Chen:2019ejq,DES:2019fny}. However, a drawback of these $H_0$ measurements is that the results from them are strongly cosmological model-dependent \citep{Qi:2022kfg}. For example, in the $w$CDM model with a constant equation-of-state parameter of dark energy $w$, the inferred $H_0$ value shifts to $81.6^{+4.9}_{-5.3} ~\rm{km\ s^{-1}\ Mpc^{-1}}$ \citep{Wong:2019kwg}.

With the distance sum rule, \citet{Collett:2019hrr} proposed a cosmological model-independent method to determine $H_0$ and $\Omega_K$ simultaneously from the precise measurements of SGLTD, in which SNe Ia are used to serve as a distance indicator. {It should be emphasized that the distance sum rule depends on the assumption of the cosmological principle that the universe is described by the homogeneous and isotropic Friedmann-Lema\^{\i}tre-Robertson-Walker (FLRW) metric on large scales. As a cornerstone of cosmology, the validity of the FLRW metric has been verified by the growing observational data of increased precision \cite{Clarkson:2007pz,Shafieloo:2009hi,Sapone:2014nna,Rasanen:2014mca,Qi:2018atg,Cao:2019kgn,Rasanen:2013swa}. In addition, this method also relies on modelling the distance function with a polynomial. By fitting to the mock data generated by different models and performing an out-of-sample error analysis based on real data, Collett et al. \citep{Collett:2019hrr} found that the typical deviation between the average value of polynomial and the real underlying value is less than $10\%$ of the statistical uncertainty. Moreover, Qi et al. \citep{Qi:2020rmm} have demonstrated that a third-order polynomial is flexible enough to fit the current data. Therefore, this cosmological model-independent method of determining $H_0$ and $\Omega_K$ is robust. Up to now,} this approach has been extended by using SGLTD combined with other distance indicators, such as the known ultraviolet versus X-ray luminosity correlation of quasars providing luminosity distance \citep{Wei:2020suh}, the angular size of compact structure in radio quasars as standard rules \citep{Qi:2020rmm}, and gravitational wave standard sirens \citep{Cao:2021zpf,Wang:2022rvf}. The results of these previous works suggest that this cosmological model-independent method is an effective way to measure $H_0$ and $\Omega_K$ in the late universe. 

%Instead of replacing different distance indicators, in this paper, we would like to explore the possibility of improving this method from another perspective. 

Actually, we are not satisfied with only using SNe Ia as a distance indicator in the SGLTD method because they are able to play a more significant role. In this paper, we consider to use SNe Ia as lensing sources in the SGLTD method, rather than only using them as a distance indicator, to show what influences of SNe Ia could bring into cosmology with the SGLTD method.

As mentioned above, the time-delay measurements originally proposed to measure $H_0$ are actually to use strongly lensed SNe. In fact, lensed SNe have several advantages over lensed quasars in accurate measurements of time delays. (i) SN is a transient source whose sharply varying light-curve shape makes it easy to measure the time delay and less prone to be strongly inflenced by microlensing effects \citep{Suyu:2020opl}. (ii) Unlike the strong contamination by quasar light that outshines everything else in the lens system, the lens galaxy could be observed clearly after an SN fades, allowing accurate modelling of lens mass distribution \citep{Suyu:2020opl}. (iii) If the lensed source is an SN Ia, the lens model degeneracies could be mitigated due to the standard intrinsic luminosities in the cases when microlensing effects are negligible  \citep{Suyu:2020opl}. (iv) For a lensed SN, the effect of microlensing time delay could be ignored \citep{Bonvin:2018lgh}, but it cannot be ignored in the case of a lensed quasar \citep{tie2018microlensing}. Finally, another advantage we wish to emphasize is that the observational duration for a lensed SN is much shorter than the decades-long observation of a lensed quasar. With so many advantages, lensed SNe are expected to be a powerful cosmological probe. 

%Since current studies on lensed SNe mainly discuss SNe Ia as the lensed sources, all lensed SNe we discuss in following refer to Type Ia supernovae as the lensed sources.

Although there are few events of lensed SNe Ia currently, the observational event number will greatly increase thanks to the ongoing and future massive surveys, such as Dark Energy Survey (DES), Zwicky Transient Facility (ZTE) \citep{bellm2018zwicky}, and Legacy Survey of Space and Time (LSST) \citep{LSST:2008ijt,Huber:2019ljb}. According to some estimates \citep{Oguri:2010ns,Collett:2015roa,Goldstein:2018bue,Wojtak:2019hsc}, the LSST will observe hundreds of lensed SNe Ia in a 10-year observation. With the upcoming boom in lensed SNe, the HOLISMOKES (Highly Optimised Lensing Investigations of Supernovae, Microlensing Objects, and Kinematics of Ellipticals and Spirals) project was set up to find and measure strongly lensed SNe in current/future surveys, for which we refer the reader to Refs.~\cite{Suyu:2020opl,Canameras:2020ymx,Huber:2020dxc,Bayer:2021ugw,Canameras:2021uco,Huber:2021iug}. As HOLISMOKES demonstrated \cite{Suyu:2020opl}, the possibility of using strongly lensed SNe Ia as an accurate cosmological probe will soon become a reality \cite{Lochner:2021zpr, Abell:2009aa, Marshall:2017wph, Goldstein:2018bue, Huber:2019ljb}.

In view of the broad prospect for the time-delay measurements in the upcoming LSST era, in this paper, we will use lensed SNe Ia as an accurate probe to promote the cosmological model-independent measurements of $H_{0}$ and $\Omega_{K}$. Here we adopt the latest Pantheon SNe Ia sample as the distance indicator to determine the distances from the observer to the lens and source in a SGL system. We simulate a sample of lensed SNe Ia based on the LSST survey and forecast what precision can be achieved for the constraints on $H_0$ and $\Omega_K$. For comparison, we will also simulate a sample of lensed quasars in the ear of LSST, and implement the same constraints.

\section{Methodology}

The universe is homogeneous and isotropic on large scales, which is described by the FLRW metric,
\begin{align}
d s^{2}=-c^{2} d t^{2}+a^{2}(t)\left(\frac{d r^{2}}{1-K r^{2}}+r^{2} d \Omega^{2}\right),\label{1}
\end{align}
where $c$ is the speed of light and $a(t)$ is the scale factor. The constant $K$ represents the spatial curvature, which is related to the curvature parameter $\Omega_{K}$ and the Hubble constant $H_{0}$ as $\Omega_{K}=-Kc^{2}/a_{0}^{2}H_{0}^{2}$.

For an SGL system, the dimensionless comoving distance between the lens at redshift $z_{l}$ and the source at redshift $z_{s}$ can be expressed as
\begin{align}
d\left(z_{l}, z_{s}\right)=\frac{1}{\sqrt{\left|\Omega_{K}\right|}} \operatorname{sinn}\left(\sqrt{\left|\Omega_{K}\right|} \int_{z_{l}}^{z_{s}} \frac{H_{0}}{H(z)} d z\right),\label{2}
\end{align}
where
\begin{align}
\operatorname{sinn}(x)= \begin{cases}\sin (x), & \Omega_{K}<0, \\ x, & \Omega_{K}=0, \\ \sinh (x), & \Omega_{K}>0.\end{cases}\label{3}
\end{align}
For convenience, we denote $d_{l} \equiv d\left(0, z_{l}\right)$, $d_{s} \equiv d\left(0, z_{s}\right)$ and $d_{l s} \equiv d\left(z_{l}, z_{s}\right)$. The three dimensionless comoving distances are connected via the distance sum rule \cite{Rasanen:2014mca, Xia:2016dgk, Li:2018hyr, Liao:2019hfl, Liao:2019qoc, Qi:2018atg, Qi:2020rmm, Wang:2019yob, Wang:2020dbt, Zhou:2019vou}:
\begin{align}
d_{l s}=d_{s} \sqrt{1+\Omega_{K} d_{l}^{2}}-d_{l} \sqrt{1+\Omega_{K} d_{s}^{2}}.\label{4}
\end{align}
Furthermore, Equation (\ref{4}) can be rewritten as
\begin{align}
\frac{d_{l} d_{s}}{d_{l s}}=\frac{1}{\sqrt{1 / d_{l}^{2}+\Omega_{K}}}-\frac{1}{\sqrt{1 / d_{s}^{2}+\Omega_{K}}}.\label{5}
\end{align}

If the arrival times of two lensed images are marked as $t_{i}$ and $t_{j}$, respectively, the time delay $\Delta t_{i,j}$ is related to the time-delay distance $D_{\Delta t}$ and the Fermat potential difference $\Delta \phi_{i,j}$ as
\begin{align}
\Delta t_{i, j}=\frac{\left(1+z_{l}\right) D_{\Delta t}}{c} \Delta \phi_{i, j}.\label{6}
\end{align}
Here $\Delta \phi_{i,j}$ is given by
\begin{align}
\phi_{i,j}=\left[\frac{\left(\theta_{i}-\beta\right)^{2}}{2}-\psi\left(\theta_{i}\right)-\frac{\left(\theta_{j}-\beta\right)^{2}}{2}+\psi\left(\theta_{j}\right)\right],\label{7} \end{align}
where $\theta_{i}$ and $\theta_{j}$ are {the angular positions} of two images, respectively, $\beta$ represents the angular position of source and $\psi$ is the two-dimensional lens potential which depends on the mass distribution of the lens. The time-delay distance is composed of three angular diameter distances. Based on the relationship between the dimensionless comoving distance and the angular diameter distance $d\left(z_{l}, z_{s}\right) \equiv\left(1+z_{s}\right) H_{0} D_{A}\left(z_{l}, z_{s}\right) / c$, the time-delay distance can be rewritten as
\begin{align}
D_{\Delta t} \equiv\left(1+z_{l}\right) \frac{D_{l}^{A} D_{s}^{A}}{D_{ls}^{A}}=\frac{c}{H_{0}} \frac{d_{l} d_{s}}{d_{l s}}.\label{8}
\end{align}

It can be seen clearly from Equations (\ref{5}) and (\ref{8}) that once the dimensionless comoving distances $d_{l}$ and $d_{s}$ are obtained, the Hubble constant $H_{0}$ and the cosmic curvature $\Omega_{K}$ can be determined from the measurements of time-delay distance without any specific cosmological model.

\subsection{Time-delay distance from lensed SNe Ia}
Here, we briefly introduce the simulation of the time-delay distance measurements from lensed SNe Ia. Recent analyses revealed that several hundred lensed SNe Ia could be observed in the 10-year $z$-band search of the LSST survey \cite{LSSTDarkEnergyScience:2021ryz, Oguri:2010ns, Collett:2015roa, Goldstein:2016fez,Birrer:2021use}. However, it should be noted that the different observing strategies with different survey areas and different cumulative season lengths will have different estimates of the number of observed lensed SNe Ia \citep{Huber:2019ljb}. Therefore, in this paper, we consider different scenarios with the various well-measured lensed SNe Ia numbers of $N_{\rm SN}=20$, 50, 100, 150, and 200, respectively.

For the lens modelling, we adopt the singular isothermal ellipsoid model \citep{kormann1994isothermal,Barkana:1998qu} that is in good agreement with observations to characterize the mass distribution for all lens galaxies. For the uncertainties of time-delay distance $D_{\Delta t}$ measurements, there are three factors considered in our simulation: the time delay, the Fermat potential difference, and the mass distribution along the line of sight (LOS) to the lensing source \cite{Wong:2019kwg, Millon:2019slk, Birrer:2020tax, Ding:2021bxs}. For the measurements of time delay, although the microlensing affects all images, \citet{Goldstein:2016fez} proposed that one can use the color curves of SNe instead of the broadband light curves to extract time delay with high precision. Moreover, by fitting flux and color observations of microlensed SNe Ia with their underlying, unlensed spectral templates, \citet{Goldstein:2017bny} demonstrated that the fitting of the template to light curves yields a 4\% uncertainty for time-delay measurements due to microlensing, whereas the microlensing-induced time delay uncertainty decreases to 1\% when the template is fitted to color curves in the achromatic phase. \citet{Pierel:2019pnr} developed an open-source package, Supernova Time Delays (SNTD), to make accurate time-delay measurements of lensed SNe Ia including treatments of microlensing. By running an automated fitting algorithm on the simulated data using a variety of tools in SNTD, they found that obtaining before-peak observations of light curves improves the precision of time-delay measurements to 3\%. Therefore, in this paper, we optimistically adopt a 3\% uncertainty for the lensed SNe Ia time-delay measurements. While, for the lensed quasars, the uncertainty of time delay is about 5\% \citep{Suyu:2020opl,Chen:2019ejq,Tewes:2012zz,Vuissoz:2008qa,Bonvin:2016crt,Suyu:2016qxx}.

Since SN Ia is a transient source, a clean image of the lens galaxy could be obtained after the SN Ia fading away over time, and in the case of a lensed quasar, the lens galaxy is typically contaminated by the bright quasar light, which means that the modelling of lens mass distribution in the case of a lensed SN Ia could be improved significantly \citep{Liao:2017ioi,Qi:2018atg}. Recently, \citet{Ding:2021bxs} quantitatively estimated the improvement of lens modelling and $H_0$ inference with transient sources, resulting in an improvement of the precision for lens models by a factor of 4.1, and an improvement of $H_0$ precision by a factor of 2.9. Since the time-delay distance is inversely proportional to $H_0$, $D_{\Delta t} \propto 1/H_0$, the improvement for the precision of $D_{\Delta t}$ measurements due to the transient sources could also be realized by a factor of 2.9 compared with the case of lensed quasars. The current lensed quasar constraints \citep{Chen:2019ejq,Tewes:2012zz,Vuissoz:2008qa,Bonvin:2016crt,Suyu:2016qxx} provide a 3\% lens mass modelling uncertainty for the measurements of $D_{\Delta t}$, so here we adopt a 1\% uncertainty due to the lens mass modelling for the case of lensed SNe Ia. An additional 3$\%$ uncertainty is derived from the LOS effect, which is given by current lensed quasar observations \cite{Suyu:2020opl, Liao:2014cka}. Table~\ref{Tab1} lists the relative uncertainties of factors contributing to the time-delay distance measurements for the lensed SNe Ia and lensed quasars for comparison.

{For the redshift distribution of SNe Ia as sources in the SGL systems, it has been calculated from the mock SGL catalogue constructed using a Monte Carlo technique based on the  LSST observing strategies (see Figure 5 of Ref. \citep{Oguri:2010ns}), in which we find that the maximum redshift of SNe Ia extends to about 2.2. Although the SNe Ia with higher redshifts can also be expected to be observed due to the magnification of strong lensing, high redshift systems are overall fainter and the larger photometric errors make the time delay measurements more uncertain \citep{Huber:2019ljb}. At present, the redshifts of sources for two observed lensed SNe Ia, SN Refsdal \citep{Kelly:2014mwa} and iPTF16geu \citep{Goobar:2016uuf}, are 1.49 and 0.409, respectively, which are consistent with the prediction of Oguri and Marshall \citep{Oguri:2010ns}. For our approach, this means that the existing Pantheon sample of SNe Ia could be used to calibrate almost all lensed SNe Ia observed in the LSST era.}

Finally, we generate a sample of time-delay distances based on the Orguri and Marshall catalog \citep{Oguri:2010ns} in the flat $\Lambda$CDM model with the matter density $\Omega_{m}=0.3$ and the Hubble constant $H_{0}=73.2$ $\rm km\ s^{-1}\ Mpc^{-1}$.

\begin{table}
\renewcommand\arraystretch{1.5}
\caption{Uncertainties of three sources contributing to the uncertainty of time-delay distance measurements.}
\label{Tab1}
\centering
\normalsize
\begin{tabular}{p{2.5cm} p{2cm}<{\centering} p{2cm}<{\centering}  p{2cm}<{\centering}}
%\bottomrule[1pt]
\hline\hline
       SGL source                     & $\delta\Delta t$         & $\delta\Delta \psi$            & $\delta\rm LOS$         \\
\hline
Lensed SNe Ia              & 3\%                  &  1\%                              &  3\%           \\
Lensed quasars             & 5\%                &  3\%                               &  3\%                \\
%\bottomrule[1pt]
\hline\hline
\end{tabular}
\end{table}

\subsection{Distance calibration by SNe Ia}

In this paper, we use the latest SNe Ia data from the Pantheon sample \cite{Pan-STARRS1:2017jku} as a distance indicator to calibrate the dimensionless comoving distances $d_{l}$ and $d_{s}$ of SGL systems. The sample consists of 1048 SNe Ia data covering the redshift range $0.001 < z < 2.3$. 

As ``standard candles'', the SNe Ia observational data are connected with the distance modulus through the SALT2 light curve fitter \citep{SNLS:2010pgl}:
\begin{align}
\mu=m_{B}+\alpha \cdot X_{1}-\beta \cdot \mathcal{C}-M_{B},\label{10}
\end{align}
where $m_{B}$ is the rest frame B-band peak magnitude, $X_{1}$ and $\mathcal{C}$ represent the time stretch of light curve and the supernova color at maximum brightness, respectively, and $M_{B}$ is the absolute B-band magnitude. The stretch-luminosity parameter $\alpha$ and the color-luminosity parameter $\beta$ are two nuisance parameters in the distance estimation. To dodge this problem, $\alpha$ and $\beta$ could be calibrated to zero by the BEAMS with Bias Corrections method \cite{Kessler:2016uwi}. Then the observed distance modulus can be simply expressed as
\begin{align}
\mu=m_{B}-M_{B}.\label{11}
\end{align}

The luminosity distance $D_{L}$ of an SN Ia is related to the distance modulus as
\begin{align}
\mu=5 \log_{10} \left(D_{L}\right)+25.\label{12}
\end{align}
We can see that once the value of $M_{B}$ is determined, the luminosity distance of an SN Ia can be obtained. In this work, we regard $M_{B}$ as a free parameter due to the degeneracy between $M_{B}$ and $H_{0}$. Based on the relation between the comoving dimensionless distance and the luminosity distance, we have
\begin{align}
d(z)=\frac{H_{0} D_{L}(z)}{c(1+z)}.\label{13}
\end{align}

The difficulty of using SNe Ia data to calibrate the distances of SGL systems is that the redshifts of the two data sets cannot be one-to-one correspondence. In this paper, we establish a continuous distance-redshift function using a polynomial fit to treat this issue. The theoretical $d(z)$ used in Equations (\ref{5}) and (\ref{8}) is assumed following a third-order polynomial:
\begin{align}
d(z)=z+a_{1} z^{2}+a_{2} z^{3}.\label{14}
\end{align}
As long as $d(z)$ is more flexible than a second-order polynomial, there is not much of difference \cite{Rasanen:2014mca, Liao:2017yeq}. So the third-order polynomial we adopted is flexible enough to fit the distance data  \citep{Qi:2020rmm}.

Finally, we constrain the cosmological parameters using the {\tt emcee} Python module \cite{Foreman-Mackey:2012any} based on the Markov Chain Monte Carlo analysis. The final likelihood function is $\mathcal{L} \propto e^{-\chi^{2} / 2}$, and the $\chi^2$ function is defined as
\begin{align}
\chi^{2}=\sum_{i=1}^{N_{\text{len}}} \left(\frac{D_{\Delta t,i}^{\mathrm{th}}-D_{\Delta t,i}^{\mathrm{obs}}}{\delta D_{\Delta t,i}^{\mathrm{obs}}}\right)^{2}+\sum_{i=1}^{1048} \left(\frac{d(z_{i})^{\mathrm{th}}-d(z_{i})^{\mathrm{obs}}}{\delta d_{i}^{\mathrm{obs}}}\right)^{2},\label{15}
\end{align}
where ``obs'' represents the observation, and ``th'' represents the theoretical value derived from distance sum rule. {Here, $d(z)^{\rm{obs}}_{i}$ is the dimensionless comoving distance derived from the Pantheon sample of SNe Ia via Equations (\ref{11})--(\ref{13}), and $\delta d^{\rm{obs}}_{i}$ is the corresponding error.}
The whole free parameters include $H_{0}$, $\Omega_{K}$, $a_{1}$, $a_{2}$ and $M_{B}$. In this work, we only focus on the constraints on $H_{0}$ and $\Omega_{K}$.

\section{Results and discussion}

\begin{table*}  
\renewcommand\arraystretch{1.5}
\caption{1$\sigma$ uncertainty results of measuring $H_0$ and $\Omega_K$ for various event numbers of lensed SNe Ia. Here $H_{0}$ is in units of $\rm km\ s^{-1}\ Mpc^{-1}$.}\label{Tab2}
\begin{center}
\normalsize
\setlength{\tabcolsep}{3mm}{
\begin{tabular}{p{4cm}<{\centering} p{1.8cm}<{\centering} p{1.8cm}<{\centering} p{1.8cm}<{\centering} p{1.8cm}<{\centering} p{1.8cm}<{\centering}}
%\hline
%\bottomrule[1pt]
\hline\hline
Parameter error & $20$ SN &$50$ SN & $100$ SN & $150$ SN & $200$ SN \\ 
\hline
$\Delta H_0$ (free $\Omega_K$) & 1.52 & 0.96 & 0.72 & 0.60 & 0.52\\
$\Delta H_0$ (fixed $\Omega_K$) & 0.85 & 0.59 & 0.45 & 0.38 & 0.33\\
%\hline
$\Delta \Omega_K$ (free $H_0$)  & 0.158 & 0.108 & 0.093 & 0.084 & 0.078 \\
$\Delta \Omega_K$ (fixed $H_0$)  & 0.087 & 0.069 & 0.060 & 0.055 & 0.052 \\
\hline\hline
%\bottomrule[1pt]
\end{tabular}}
\end{center}
\end{table*}

\begin{table*}  
\renewcommand\arraystretch{1.5}
\caption{1$\sigma$ uncertainty results of measuring $H_0$ and $\Omega_K$ for various event numbers of lensed quasars. Here $H_{0}$ is in units of $\rm km\ s^{-1}\ Mpc^{-1}$.}\label{Tab3}
\begin{center}
\normalsize
\setlength{\tabcolsep}{3mm}{
\begin{tabular}{p{4cm}<{\centering} p{1.8cm}<{\centering} p{1.8cm}<{\centering} p{1.8cm}<{\centering} p{1.8cm}<{\centering} p{1.8cm}<{\centering}}
\hline\hline
%\bottomrule[1pt]
Parameter error & $50$ QSO &$100$ QSO & $200$ QSO & $300$ QSO & $400$ QSO \\ 
\hline
$\Delta H_0$ (free $\Omega_K$) & 1.56 & 1.25 & 0.80 & 0.67 & 0.59\\
$\Delta H_0$ (fixed $\Omega_K$) & 0.80 & 0.64 & 0.47 & 0.42 & 0.36\\
%\hline
$\Delta \Omega_K$ (free $H_0$) & 0.168 & 0.122 & 0.096 & 0.088 & 0.085 \\
$\Delta \Omega_K$ (fixed $H_0$)  & 0.087 & 0.069 & 0.061 & 0.054 & 0.052 \\
\hline\hline
%\bottomrule[1pt]
\end{tabular}}
\end{center}
\end{table*}

\begin{figure}
\begin{center}
\includegraphics[scale=0.62]{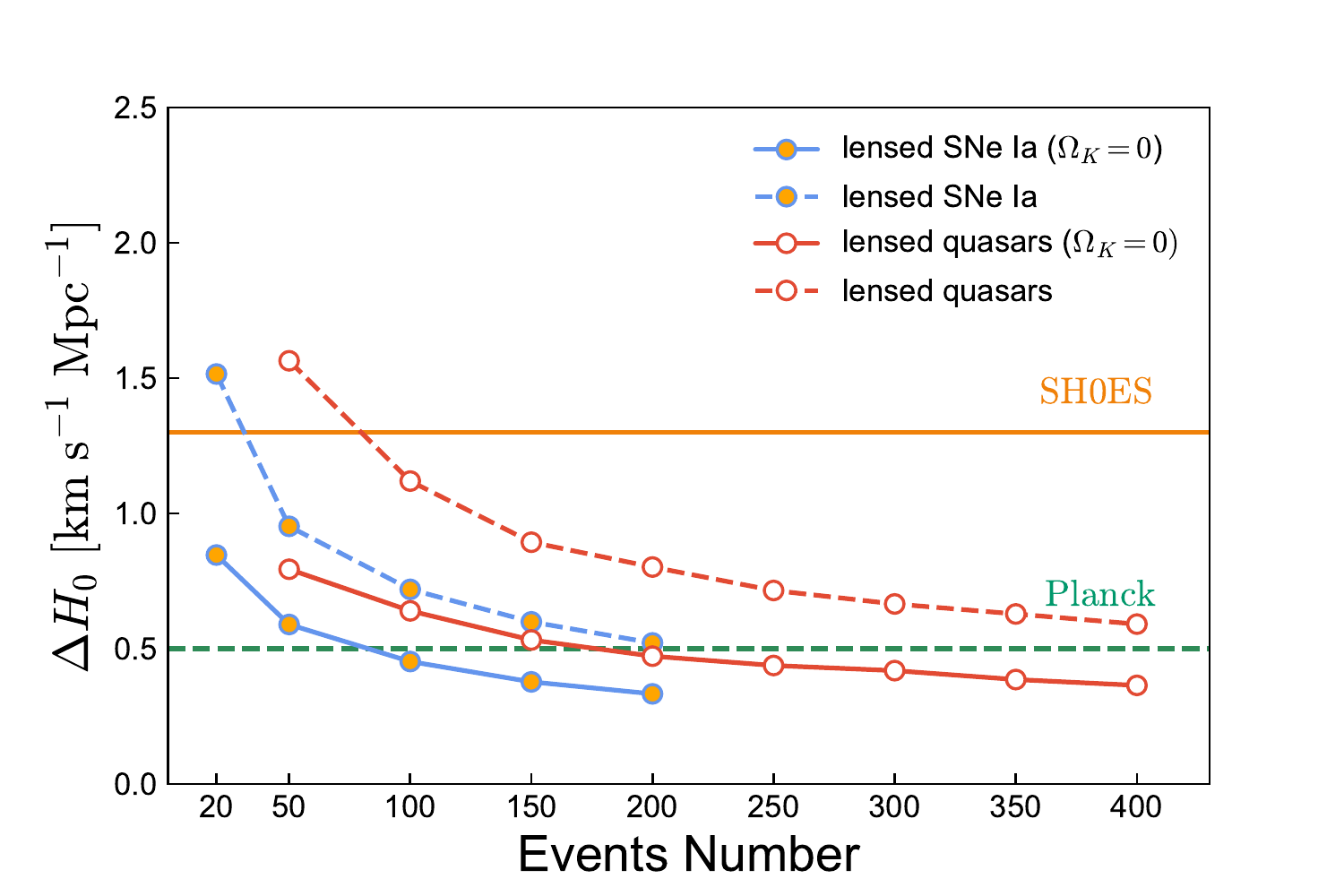}
\caption{The 1$\sigma$ errors of $H_{0}$ for different numbers of lensed SNe Ia and lensed quasars. The blue and red lines represent the results from lensed SNe Ia and lensed quasars, respectively. The dotted polyline denotes the uncertainties of $H_0$ obtained in the case of treating both $H_0$ and $\Omega_K$ as free parameters, and the solid polyline denotes $\Delta H_0$ obtained with a prior of $\Omega_K=0$. The orange and green horizontal lines represent the measured uncertainties of $H_{0}$ by SH0ES and \emph{Planck} collaborations, respectively. Here $H_0$ is in units of $\rm km\ s^{-1}\ Mpc^{-1}$. \label{Fig1}}
\end{center}
\end{figure}

\begin{figure}
\begin{center}
\includegraphics[scale=0.62]{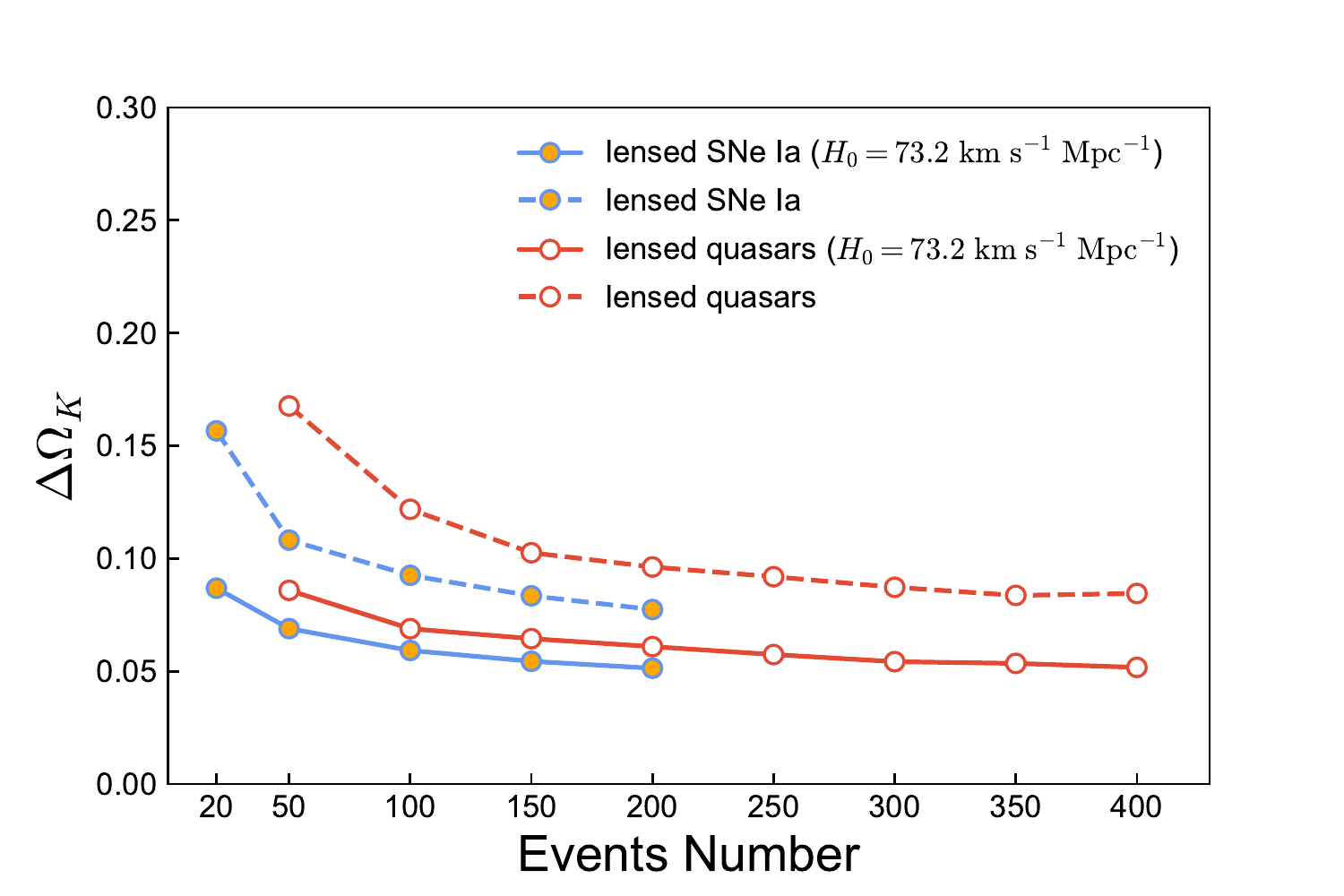}
\caption{The 1$\sigma$ uncertainties of $\Omega_K$ for different numbers of lensed SNe Ia and lensed quasars. The blue and red lines represent the results from lensed SNe Ia and lensed quasars, respectively. The dotted polyline denotes the uncertainties of $\Omega_K$ obtained in the case of treating both $H_0$ and $\Omega_K$ as free parameters, and the solid polyline denotes $\Delta \Omega_K$ obtained with a prior of $H_0=73.2 ~\rm km\ s^{-1}\ Mpc^{-1}$.\label{Fig2}}
\end{center}
\end{figure}

\begin{figure}
\begin{center}
\includegraphics[scale=0.26]{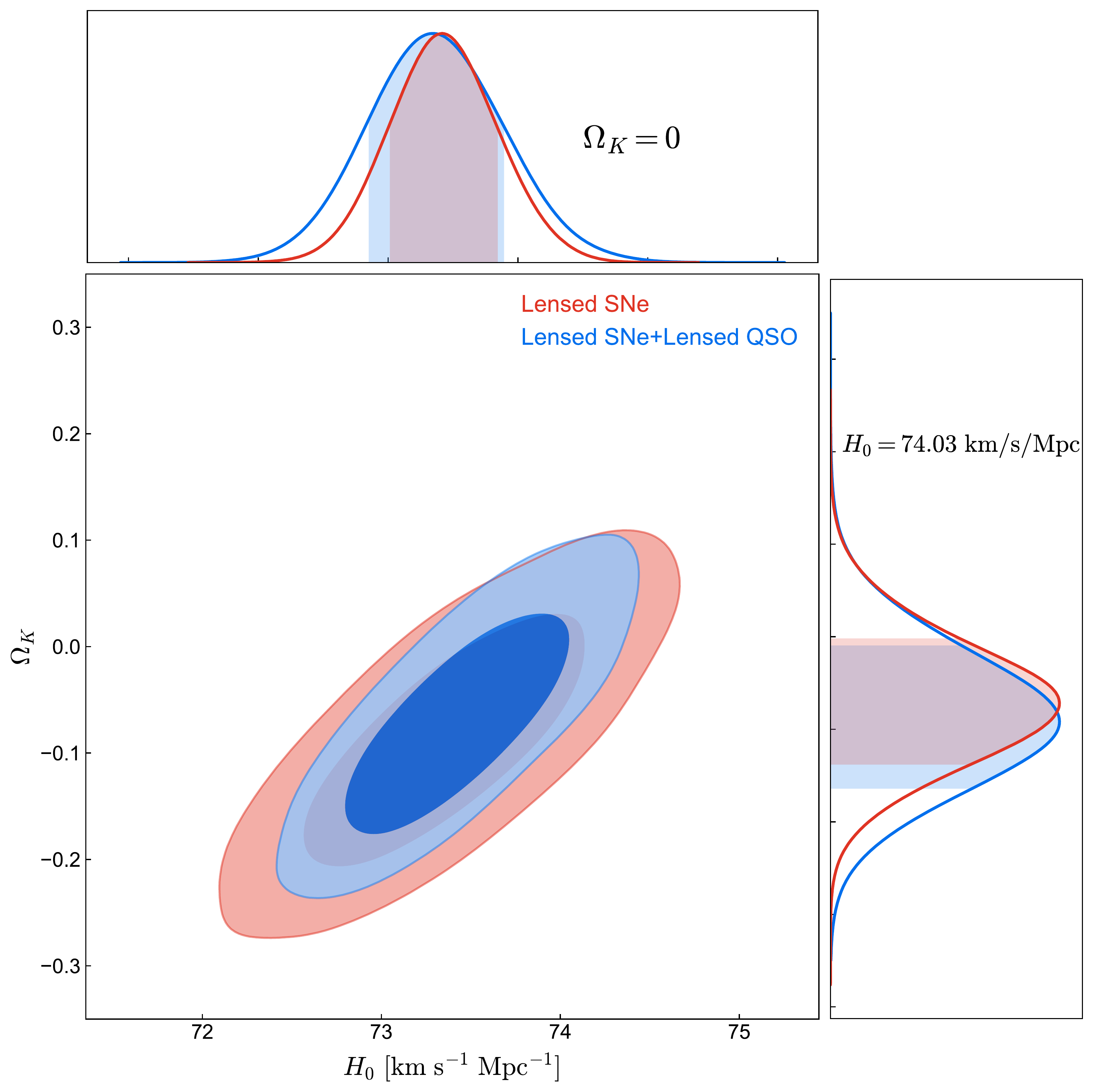}
\caption{1D and 2D marginalized probability distributions of $H_{0}$ and $\Omega_{K}$ constrained from 200 lensed SNe Ia and 400 lensed quasars + 200 lensed SNe Ia. \label{Fig3}}
\end{center}
\end{figure}

In this section, we report our constraint results in detail and make some discussions. In Section \ref{sec_SN}, we will show the constraints on $H_{0}$ and $\Omega_{K}$ using time delay from lensed SNe Ia and the distance calibration from Pantheon SNe Ia. In Section \ref{sec_QSO}, we will make a comparison with the case of lensed quasars for the capabilities of constraining cosmological parameters. The constraint results are displayed in Figures \ref{Fig1}--\ref{Fig3} and summarized in Tables \ref{Tab2} and \ref{Tab3}. It should be noted that we use $\Delta \xi$ to represent the error of a parameter $\xi$.

\subsection{Constraints on cosmological parameters from lensed SNe Ia}\label{sec_SN}

Here, we present the constraint errors of $H_{0}$ and $\Omega_{K}$ for various detected numbers of lensed SNe Ia in Figures \ref{Fig1} and \ref{Fig2} and Table \ref{Tab2}. Firstly, we focus on the constraints on $H_0$ as indicated by the blue lines in Figure \ref{Fig1}. The blue dotted polyline represents the results obtained when both $H_0$ and $\Omega_K$ are free parameters. We can see that even in the most conservative case ($N_{\text{SN}}=20$), the result of $\Delta H_0=1.52$ $\rm km\ s^{-1}\ Mpc^{-1}$ can be obtained using this cosmological model-independent method, which is comparable to the result of $\Delta H_0=1.3$ $\rm km\ s^{-1}\ Mpc^{-1}$ given by the SH0ES collaboration \citep{Riess:2020fzl}. If 200 lensed SNe Ia could be observed, the constraint on $H_{0}$ will be improved to $\Delta H_0=0.52~\rm km\ s^{-1}\ Mpc^{-1}$, which is comparable with (and actually even slightly better than) the result of $\Delta H_0 =0.54$ $\rm km\ s^{-1}\ Mpc^{-1}$ from \emph{Planck} 2018 TT,TE,EE+lowE+lensing data \cite{Aghanim:2018eyx}. Such a result meets the standard of precision cosmology.

If we consider a flat universe (with fixed $\Omega_K=0$), as shown by the blue solid polyline in Figure \ref{Fig1}, the constraint on $H_0$ will be improved significantly. In this case, the constraint from only 20 lensed SNe Ia, $\Delta H_0=0.85 ~\rm km\ s^{-1}\ Mpc^{-1}$, could exceed the result of $\Delta H_0=1.3 ~\rm km\ s^{-1}\ Mpc^{-1}$ by the SH0ES collaboration \citep{Riess:2020fzl}. Only 100 lensed SNe Ia are required to produce results better than those obtained by \emph{Planck} 2018 TT, TE, EE+lowE+lensing data. In the most optimistic scenario ($N_{\text{SN}}=200$), we can get a result of $\Delta H_0=0.33$ $\rm km\ s^{-1}\ Mpc^{-1}$. All these results demonstrate that the lensed SNe Ia are competent as a precise cosmological probe, and using the time-delay measurements of lensed SNe Ia could provide an effective method of measuring $H_0$, which is precise enough to address the Hubble tension issue. 

%as a third-party criterion to explore the $H_0$ tension problem.

In Figure \ref{Fig2}, we show the constraint errors on $\Omega_K$. The blue dotted polyline represents the constraint errors for different numbers of lensed SNe Ia when both $H_0$ and $\Omega_K$ are free parameters, and the blue solid polyline represents the results when only $\Omega_K$ is a free parameter ($H_0$ is fixed to $H_0=73.2~\rm km\ s^{-1}\ Mpc^{-1}$). We find that with and without the prior of $H_0=73.2~\rm km\ s^{-1}\ Mpc^{-1}$, the constraint on $\Omega_K$ changes significantly. This actually means that there is a degeneracy between $\Omega_K$ and $H_0$, which can also be verified from the constraint on $H_0$ shown in Figure \ref{Fig1}. {One interesting point is that the constraint on $\Omega_K$ is not improved significantly as the number of lensed SNe Ia increases, especially after the number reaches 100. In principle, if there is no systematic error but only statistical error, the constraint precision of parameters will by improved by a factor of $\sqrt{N}$ as the number of observational data is increased by a factor of $N$. However, our results show that the influence of systematic errors cannot be ignored, and actually the systematic errors will dominate over statistical errors when the number of the SGL data is increased to a certain extent.} Here, we report the results we obtained. In the most conservative estimate ($N_{\text{SN}}=20$), the results for $\Omega_K$ are $\Delta \Omega_K=0.158$ and $\Delta \Omega_K=0.087$ corresponding to the cases with two free parameters (free $H_0$ and $\Omega_K$) and with one free parameter (only $\Omega_K$ as a free parameter), respectively. In the optimistic scenario ($N_{\text{SN}}=200$), we get $\Delta \Omega_K=0.078$ and $\Delta \Omega_K=0.052$, respectively. Although the constraints on $\Omega_K$ here are not as good as the result with a $1\sigma$ error of 0.002 obtained from \emph{Planck} 2018 TT, TE, EE+lowE+lensing+BAO data \citep{Aghanim:2018eyx}, it must be emphasized that our constraints are independent of any cosmological models, which will be helpful in solving cosmological tension problem concerning the cosmic curvature in the future.

\subsection{Comparison with constraints from lensed quasars}\label{sec_QSO}
As mentioned above, lensed SNe Ia have several advantages over lensed quasars in accurate and precise measurements of time delays. However, in cosmological applications, lensed quasars have an advantage over lensed SNe Ia in that they have a larger sample size. In this subsection, we also investigate the capability of constraining cosmological parameters with lensed quasars and make a comparison with the case of lensed SNe Ia. According to the forecasts in Ref.~\citep{Oguri:2010ns}, LSST will find about 400 lensed quasars with well-measured time delays. We simulate lensed quasar samples in different scenarios ($N_{\rm QSO}=50$, 100, 200, 300, and 400, respectively). Based on the methods and algorithms of current surveys, the uncertainty of measuring time-delay distances from lensed quasars is assumed to be $6.6\%$ \cite{Suyu:2020opl}. The uncertainties of factors contributing to the final uncertainty of time-delay distances are summarized in Table \ref{Tab1}.

In the case of both $H_0$ and $\Omega_K$ being free parameters, the constraint errors of $H_0$ for various numbers of lensed quasars are represented by the red dotted polyline in Figure \ref{Fig1}. The red solid polyline denotes the results in a flat universe, i.e., with fixed $\Omega_K=0$. The constraint results are summarized in Table \ref{Tab3}. We can clearly see that with the same data size, the constraints on $H_0$ from lensed SNe Ia are much better than those from the lensed quasars. Moreover, after the number of lensed quasars exceeds 150, the constraints on $H_0$ are improved very little as the number increases. Even if the event number of lensed quasars could reach the most optimistic 400, the constraint on $H_0$ from them is less precise than that from 200 lensed SNe Ia. Nevertheless, the strongly lensed quasars are still an effective late-universe probe. 150 lensed quasars data could achieve a constraint on $H_0$ of $\Delta H_0= 0.5$ $\rm km\ s^{-1}\ Mpc^{-1}$, which is comparable with the result from the {\it Planck} 2018 data \citep{Aghanim:2018eyx}.

With and without the prior of $H_{0}=73.2$ $\rm km\ s^{-1}\ Mpc^{-1}$, we obtain the constraint errors of $\Omega_K$ represented by the red solid polyline and red dotted polyline, respectively, in Figure \ref{Fig2}. In the most optimistic case, with the prior of $H_0$, we obtain the tightest constraint $\Delta \Omega_K=0.052$ from 400 lensed quasars, which is the same as the result of $\Delta \Omega_K=0.052$ from 200 lensed SNe Ia. We can see that even though the detectable number of lensed SNe Ia is small ($N_{\rm SN}=200$), the constraints on $H_0$ and $\Omega_K$ from lensed SNe Ia are comparable with those obtained by the much larger sample of lensed quasars ($N_{\rm QSO}=400$).

In the LSST era, many lensed SNe Ia and lensed quasars will be observed simultaneously. Here, in the most optimistic case, we can make a prediction for the constraints on $H_0$ and $\Omega_K$ from the combination of 200 lensed SNe Ia and 400 lensed quasars. In Figure \ref{Fig3}, the 2D contours represent the constraint results from lensed SNe Ia and the combination of lensed SNe Ia and lensed quasars. The 1D probability distribution of $H_0$ is obtained with the prior of $\Omega_K=0$, and the 1D probability distribution of $\Omega_K$ is obtained with the prior of $H_{0}=73.2~\rm km\ s^{-1}\ Mpc^{-1}$. Comparing with the result from 200 lensed SNe Ia alone, we can clearly see that combining 400 lensed quasars does not significantly improve the constraints on $H_0$ and $\Omega_K$. Here, we report that the limits of constraints on $H_0$ and $\Omega_K$ in the near future by using our method are $\Delta H_0 =0.26~\rm km\ s^{-1}\ Mpc^{-1}$ and $\Delta \Omega_K=0.044$ from the combination of 200 lensed SNe Ia and 400 lensed quasars.

\section{Conclusion}

Considering the inconsistencies in measuring some key cosmological parameters (such as the Hubble constant and cosmic curvature), reflecting the conflict between the measurements of early and late universe, one of the most important missions in modern cosmology is to develop novel, precise cosmological probes to re-examine the late-universe constrains on related parameters. SGLTD from lensed quasars as an effective cosmological probe has provided a powerful tool to measure $H_0$. However, a drawback of this $H_0$ measurement is that it is strongly cosmological model-dependent.

In this paper, we propose a scheme of using the strongly lensed SNe Ia to improve the measurements on the Hubble constant and cosmic curvature. Firstly, the distance sum rule in SGL provides a cosmological model-independent method to determine the Hubble constant and cosmic curvature simultaneously, by which we present the constraints on $H_0$ and $\Omega_K$ from the SGLTD measurements in the era of LSST. Secondly, the lensed SNe Ia with several advantages over lensed quasars in accurate measurements of time-delay distance enables us to expect the tighter constraints on $H_0$ and $\Omega_K$ using such a method. We generate a series of mock samples of lensed SNe Ia based on the LSST survey. We find that the constraint of $\Delta H_0=0.85~\rm km\ s^{-1}\ Mpc^{-1}$ from 20 lensed SNe Ia could be achieved in a flat universe ($\Omega_K=0$), which is better than the result given by the SH0ES collaboration, and 100 lensed SNe Ia can yield a constraint better than the result from \emph{Planck} 2018 TT, TE, EE+lowE+lensing data. For the constraint on $\Omega_K$, with the prior of $H_{0}=73.2$ $\rm km\ s^{-1}\ Mpc^{-1}$, we obtain the constraint result of $0.052 \leq \Delta \Omega_K \leq 0.087$ with various numbers of lensed SNe Ia, which is not as good as the result from the {\it Planck} 2018 data, but it is a cosmological model-independent measurement in the late universe.

We also make a detailed comparison for lensed SNe Ia and lensed quasars in the upcoming LSST era. We find that compared with the same number of lensed quasars, the constraints on cosmological parameters inferred from the lensed SNe Ia are improved greatly. Nonetheless, lensed quasars still are an undeniably useful cosmological probe. In a flat universe, 50 lensed quasars could yield a tighter constraint on $H_0$ than that measured by the SH0ES collaboration, and the precision of $H_0$ from 200 lensed quasars could be comparable with the result from \emph{Planck} 2018 TT, TE, EE+lowE+lensing data. Finally, when combining 400 lensed quasar data and 200 lensed SN Ia data, the constraints are improved to $\Delta H_0=0.26$ $\rm km\ s^{-1}\ Mpc^{-1}$ and $\Delta\Omega_K=0.044$, indicating that the combination of lensed SNe Ia and lensed quasars could give tighter constraints.

In summary, lensed SNe Ia are one of the most promising late-universe probes. Using the distance sum rule, lensed SNe Ia could provide cosmological model-independent constraints on both $H_{0}$ and $\Omega_{K}$ in the upcoming LSST survey. Meanwhile, a large number of lensed quasars will also be observed, which will provide precise measurements of time delays. In the forthcoming LSST era, a tremendous increase of observed SN Ia sample will make significant improvements for constraints on $H_0$ and $\Omega_K$. Moreover, in the future, a large number of gravitational wave (GW) standard sirens with the capability of providing absolute luminosity distances will also be detected, which also could be a reliable late-universe cosmological probe. Combining GW with SGLTD has been expected to make an important contribution in measuring $H_0$ and $\Omega_K$ in the late universe \citep{Cao:2021zpf,Wang:2022rvf}. All of these are expected to place tight constraints on cosmological parameters in the late universe, and bring new opportunities in resolving the tensions between the early-universe and late-universe measurements.

\section*{Acknowledgments}
We would like to thank Kai Liao and Xu-Heng Ding for helpful discussions. This work was supported by the National Natural Science Foundation of China (Grants Nos. 11975072, 11835009, 11875102, and 11690021), the Liaoning Revitalization Talents Program (Grant No. XLYC1905011), the Fundamental Research Funds for the Central Universities (Grant Nos. N2005030 and N2105014), the National 111 Project of China (Grant No. B16009), and the science research grants from the China Manned Space Project (Grant No. CMS-CSST- 2021-B01).

\bibliography{lens_sn}

\end{document}